\documentclass[a4paper,11pt]{article}
\usepackage[a4paper, total={6in, 10in}]{geometry}
\usepackage[table,xcdraw]{xcolor}
\usepackage{subcaption}
\usepackage{amssymb}
\usepackage{rotating}
\usepackage{amsfonts}
\usepackage[official]{eurosym}
\usepackage{stackrel}
\usepackage{caption}
\usepackage{graphicx}
\usepackage{xspace}
\usepackage{tikz}
\usepackage{mathtools}
\usepackage{url}
\usepackage{authblk}
\usepackage{amsthm}
\usepackage{bm}
\usepackage[title]{appendix}
\usepackage{bbold}
\usepackage{pdfpages}
\usepackage{csquotes}
\usepackage{booktabs}
\usepackage{float}
\usepackage[english]{babel}

\theoremstyle{definition}

\mathtoolsset{showonlyrefs,showmanualtags}

\usepackage[
    backend=biber,
    style=authoryear,
  ]{biblatex}

\usepackage{hyperref}

\hypersetup{
	colorlinks=true,
	urlcolor=blue,
	citecolor=blue,
	linkcolor=blue,
        }

\DeclareCiteCommand{\textcite}
  {\boolfalse{cbx:parens}}
  {\usebibmacro{citeindex}%
   \printtext[bibhyperref]{\usebibmacro{textcite}}}
  {\ifbool{cbx:parens}
     {\bibcloseparen\global\boolfalse{cbx:parens}}
     {}%
   \multicitedelim}
  {\usebibmacro{textcite:postnote}}

\definecolor{DarkGreen}{rgb}{0,0.4,0}
\definecolor{cobalt}{rgb}{0.0, 0.28, 0.67}

\newcommand{\ts}{\textsuperscript}
\newcommand{\rev}{``}

\title{Who is in equilibrium?}
\author{Valerio Astuti}

\date{}

\begin{document}

\maketitle

\begin{abstract}
In order to describe the properties of the observed distribution of wealth in a population, most economic models rely on the existence of an asymptotic equilibrium state. 
In addition, the process generating the equilibrium distribution is usually assumed to be ergodic, with a finite asymptotic average and bounded inequality.
Here we show, using data from Bank of Italy's Survey on Household Income and Wealth and Forbes Italian billionaires lists, that the last hypothesis is not justified in Italy.
We find that, even if an equilibrium asymptotic distribution exists, the average wealth has no finite asymptotic value. As a consequence we find that - without changes in the parameters of the wealth evolution process - wealth inequality is bound to diverge with time. 
In addition we evaluate the equilibration time of the evolution process when its parameters are chosen in order to admit both an equilibrium distribution and a finite equilibrium average wealth. 
Even when both the equilibrium hypotheses are satisfied, we find equilibration times much longer than the typical time span between economic shocks. 
\end{abstract}

\section{Introduction}
Wealth inequality is steadily increasing from at least 30 years, both in Italy (\textcite{acciari2020concentration}) and in many other parts of the world (\textcite{piketty2014capital, chancel2022world}).
This fact brought the increase in inequality into the spotlight of recent academic research, with many papers trying to pinpoint the causes of such a sweeping trend.
The most common approach to the problem in the economic literature consists in the introduction of micro-founded models to describe the evolution of wealth, used to study the influence of various elements of the evolution process on the growth of inequality.
While the derivation of the properties of wealth distribution from simple assumptions on the wealth growth process dates back at least to \textcite{champernowne1953model} and \textcite{kesten1973random}, the most recent strand of economic literature on the subject can be traced back to \textcite{bewley1986stationary, huggett1993risk} and \textcite{aiyagari1994uninsured}.
There the authors emphasize the importance of uncertainty regarding the future evolution of wealth in influencing consumption choices of economic agents. 
This choices, in turn, determine the properties of the wealth distribution and wealth inequality. 
More recently, the weights of many contributing elements to wealth inequality have been studied: labor incomes (\textcite{berman2016dynamics}), capital income risk (\textcite{benhabib2015wealth}), inheritance (\textcite{piketty2015wealth, nekoei2023inheritances}), heterogeneity of returns on wealth (\textcite{gabaix2016dynamics, xavier2021wealth}), consumption again (\textcite{blanchet2022uncovering}).
A fairly recent review of the approaches explored in the literature is given in \textcite{benhabib2018skewed}.

Most of the examples cited above (and most of the existing economic literature on the subject) rely on the assumption that the evolution process is compatible with the existence of an equilibrium distribution. 
In particular the usual approach consists in assuming a given form for the wealth evolution process, finding the asymptotic equilibrium distribution for this process, and comparing the properties of the equilibrium distribution with the distribution of wealth observed in the population.  
Some of the papers cited above represent notable exception to this line of analysis: in \textcite{gabaix2016dynamics} the authors study the speed of convergence of the distribution of wealth to its equilibrium state, finding that in order for the predicted speed to be compatible with the observed one an heterogeneity in expected returns on wealth has to be introduced. 
The equilibrium notion is however still needed for the definition of the initial conditions of the dynamics, and as a consequence some of the constraints connected with the equilibrium hypothesis keep influencing the dynamics. 
In \textcite{gomez2023decomposing} and \textcite{blanchet2022uncovering}, on the other hand, a fully dynamical analysis is applied, deriving some of the properties of the wealth evolution processes from the change in time observed in the wealth distribution.  
In particular in \textcite{blanchet2022uncovering} the existence of a steady-state distribution is obtained as a prediction of the model, although some assumptions on the time evolution of the parameters of the model are needed in the estimation process. 

A different line of research on the topic has always been focused on the long-term properties of the distribution of wealth, and in particular on the existence of well defined asymptotic properties. 
In \textcite{bouchaud2000wealth} the authors, starting from a simple model of wealth growth and exchange, investigate the conditions necessary to have an asymptotic bounded inequality, and the properties of wealth distribution when these conditions are not respected. 
More recently in \textcite{bouchaud2023self} the authors introduced a model of wealth exchange based on heterogeneous beliefs. This heterogeneity implies the breaking (or, in the words of the authors, \emph{quasi-breaking}) of the ergodicity of the evolution process, and the generated wealth distribution implies large, persisting values of inequality. 
Finally in \textcite{berman2020wealth} the authors tested the hypothesis of the existence of an equilibrium asymptotic distribution for the United States, finding results which do not justify such an assumption. 
Even assuming that an equilibrium distribution exists, they find that the time necessary for the wealth evolution process to converge to such a distribution is much longer than the typical interval between economic shocks. 

In this paper we assume the dynamics of wealth to be determined by the same elementary processes studied in most of the existing economic literature on the subject (see for example \textcite{aiyagari1994uninsured}, \textcite{benhabib2015wealth}, \textcite{gabaix2016dynamics}, \textcite{blanchet2022uncovering}).
Instead of inferring the properties of these processes from the observed distribution of wealth, however, we directly estimate them from micro-data available for the Italian economy.
In particular, we exploit the Bank of Italy Survey on Household Income and Wealth (SHIW), and the Forbes Italian billionaires list. 
For previous studies exploiting the SHIW to assess properties of the distribution of income and wealth, see \textcite{brandolini2006household, cannari2018wealth, brandolini2018inequality}, while some example of use of the Forbes billionaires list are available in \textcite{klass2006forbes, korom2017enduring, gomez2023decomposing}.  
Finally, in \textcite{vermeulen2018fat} the two sources are combined to quantify the effects of the far right tail of the distribution on standard inequality indices. 
We find results compatible with the ones obtained in \textcite{blanchet2022uncovering} for the bulk of the wealth distribution, but radically different in the far right tail. 
In particular, we find that the heterogeneity of the returns process is sufficient to drive wealth inequality out of equilibrium for a large portion of the parameters space of the wealth evolution process, similarly to the results in \textcite{berman2020wealth}.


\section{Wealth evolution process}
\label{sec:wealth_growth}
In most of the economic literature on wealth inequality the evolution of wealth is assumed to be driven by a combination of stochastic processes and consumption choices of agents (see \textcite{huggett1993risk, aiyagari1994uninsured, benhabib2015wealth}).
In particular, a very general form of wealth evolution in discrete time is given by:\footnote{An alternative form of this process can be given as $x_{t+1}^i = e^{r^i_{t}}\, x^i_t +  y^i_{t} - c^i_t$. The two forms are equivalent for our scopes, but the one showed in the main text is better suited to emphasize the multiplicative nature of the process.}
\begin{equation}
\label{eq:each_agent}
    x_{t+1}^i = e^{r^i_{t}}\left( x^i_t - c^i_t  \right) +  y^i_{t}
\end{equation}
where $i$ is an index denoting each agent in the population.
In this equation, $r^i_{t}$ and $y^i_{t}$ are independent stochastic processes describing returns on wealth and labor income. 
The variable $c^i_t$, on the other hand, is usually assumed to be the optimal quantity of wealth agent $i$ decides to consume in period $t$.
This is derived from an optimization problem by which the agent balances out the utility of consuming wealth in the present period and the expected utility of consuming it in subsequent periods. 
The evolution law \eqref{eq:each_agent} encompasses most of the discrete-time models used in the economic literature, and in a suitable limit it can be used to describe a continuous-time dynamics\footnote{In appendices \ref{app:general_continuous} and \ref{app:demographic} we give a brief review of the same process in the continuous time limit.}.  

It is possible to prove that if we are interested only in the aggregate properties of the wealth distribution, any growth process of the form \eqref{eq:each_agent} can be reduced to the form:
\begin{equation}
\label{eq:generic_growth}
    \hat{x}_{t+1}\left(x_t\right) = e^{\hat{r}_{t}(x_t)}\left( x_t - \hat{c}_t\left(x_t\right) \right) + \hat{y}_{t}\left( x_t \right),
\end{equation}
where now $\hat{r}_{t}(x_t)$, $\hat{c}_t\left(x_t\right)$ and $\hat{y}_{t}\left( x_t \right)$ are all stochastic processes dependent on the present wealth value $x_t$. 
The derivation is trivial: aggregating all agents with the same wealth value $x_t = x$, we can define the empirical distribution $\rho_{x, t}\left(r, c, y\right)$ of returns, consumption and labor income. This, in turn, can be used to define the three (in general dependent) random variables $\hat{r}_{t}(x)$, $\hat{c}_t(x)$ and $\hat{y}_{t}(x)$ appearing in equation \eqref{eq:generic_growth}. 
An approach similar to this was used, in continuous time, in \textcite{blanchet2022uncovering} and \textcite{gomez2023decomposing}.
It is common to assume, in equation \eqref{eq:generic_growth}, a return process $\hat{r}_{t}(x_t)$ normally distributed and independent of the value of wealth\footnote{Notable exceptions can be found in \textcite{gabaix2016dynamics} and \textcite{xavier2021wealth}, but as we will see the assumption of returns on wealth increasing with wealth will not alter our conclusions. Results supporting the hypothesis of returns asymptotically independent of wealth are given in \textcite{levy2003rich, levy2003investment}.} and time, and of the particular realizations of the functions $\hat{c}_t(x)$ and $\hat{y}_{t}(x)$. This implies for the distribution $\rho_{x, t}\left(r, c, y\right)$ to factorize into two independent distributions $\rho_{x, t}\left(r, c, y\right) = \zeta_{x, t}\left(c, y\right) \eta\left(r\right)$, where $\eta\left(r\right)$ is a normal distribution with fixed parameters. 
While not strictly necessary for our derivation, these assumptions are common in the economic literature.
In addition, we will see in section \ref{sec:forbes} that the independence hypothesis is supported by the available data in the right tail of the wealth distribution, and while the normality assumption is only approximately valid, in appendix \ref{app:conditions_equilibrium} we will verify that the corrections to this approximation do not invalidate our results. 
Hence in the following we will assume the return process $\hat{r}_{t}(x)$ to be independent of time $t$ and wealth $x$ for large values of $x$, and to be normally distributed.

We are interested in the conditions for the existence of an equilibrium distribution of wealth, and a finite equilibrium value for inequality. 
The generic evolution equation \eqref{eq:generic_growth} allows us to study the problem independently of the assumptions of any particular model (given that any model can be reduced to this form), and in turn the conclusions obtained will be valid for all models generating an evolution described by equation \eqref{eq:each_agent}.
In particular, in this setting we can exploit a set of results available for the long-term behaviour of stochastic multiplicative processes (\textcite{champernowne1953model, kesten1973random, levy1996power, sornette1997convergent, diaconis1999iterated, mirek2011heavy}).
Expressed in terms of equation \eqref{eq:generic_growth}, a necessary condition for the existence of an asymptotic equilibrium distribution of wealth is given by (see appendix \ref{app:conditions_equilibrium} for a derivation):
 \begin{equation}
 \label{eq:eq_distr}
     \lim_{x\to \infty} \mu_t\left( x \right) < 0  
 \end{equation}
 with
 \begin{equation}
     \mu_t\left( x \right) \coloneqq \mathbf{E}\left[ \hat{r}_{t}(x) + \log\left( 1 - \frac{\hat{c}_t\left(x\right)}{x} \right) \right]
 \end{equation}
In other words for an equilibrium distribution to exist in the context of the evolution process \eqref{eq:generic_growth}, it must hold that the average return for large values of wealth is less than the value of relative consumption (we assumed for the derivation of this condition small relative consumption for large values of wealth. We will see in section \ref{sec:shiw} that this assumption is fully justified).  
In other words, on average, agents holding large values of wealth must consume more than they gain from asset returns, otherwise the value of wealth is subjected to a positive multiplicative drift, and a limit distribution cannot exist.
Demographic factors can contribute to lower the value of $\mu_t\left( x \right)$; we show in appendix \ref{app:demographic} that this is roughly equivalent to decreasing the average returns by the average mortality rate in the highest-wealth groups, adjusted for inheritance. In the following analysis we take into account these demographic factors, but they are not described explicitly in the main text for the sake of readability (see appendix \ref{app:demographic}).

In addition to the existence of an asymptotic equilibrium distribution we are interested in a stronger notion of equilibrium.
Very rarely we observe the full distribution of wealth in a population; usually we are more interested in some summary statistics, like the average wealth, its dispersion, or the Gini index. 
The existence of equilibrium values for each of these statistics is related to different constraints on the wealth evolution process, hence the existence of an equilibrium distribution does not warrant the existence of an asymptotic value of, say, the average wealth. 
The equilibrium distribution associated to a wealth growth process is usually utilized in the economic literature to describe the observed properties of the wealth distribution in a given population. 
As a consequence, the equilibrium value of any inequality measure evaluated on a simulated distribution must be equal to the values observed in the real distribution.
In particular we must assume the existence of a finite asymptotic average wealth and a Gini index smaller than one (assuming a bounded inequality in the population under study). 
In \textcite{benhabib2015wealth}, for example, the authors give a fully rigorous theoretical definition of the wealth growth process \eqref{eq:each_agent} assuming the existence of an equilibrium average wealth.
This condition is equivalent to (see appendix \ref{app:conditions_equilibrium} for a derivation):
 \begin{equation}
 \label{eq:eq_average}
     \lim_{x\to \infty} \mu'_t\left( x \right) < 0  
 \end{equation}
 with
 \begin{equation}
     \mu'_t\left( x \right) \coloneqq \log\left( \mathbf{E}\left[ e^{\hat{r}_{t}(x)} \right]\right) +   \mathbf{E}\left[ \log\left(1 - \frac{ \hat{c}_t\left(x\right) }{x} \right) \right]
 \end{equation}
Given the convexity of the exponential function, we have $\log\left( \mathbf{E}\left[ e^{\hat{r}_{t}(x)} \right] \right) \geq \mathbf{E}\left[ \hat{r}_{t}(x) \right]$.
In particular, for normally distributed returns, inequality \eqref{eq:eq_average} can be written as:
\begin{equation}
    \lim_{x\to \infty} \left\{  \mathbf{E}\left[ \hat{r}_{t}(x) \right] + \frac{1}{2}\,\mathbf{Var}\left[ \hat{r}_{t}(x) \right] +  \mathbf{E}\left[\log\left(1 - \frac{ \hat{c}_t\left(x\right) }{x} \right)\right]  \right\} < 0 
\end{equation}
For a non-trivial stochastic process, having variance $\mathbf{Var}\left[ \hat{r}_{t}(x) \right] > 0$, the last condition can be much stronger than \eqref{eq:eq_distr}, as we will see in section \ref{sec:analysis_equilibrium}.
The average return now must be smaller then the sum of the average relative consumption and the \emph{negative} half variance of the return itself.  
In particular, we will see that if demographic factors are not taken into account this implies a negative average return in order to have an asymptotic finite value for the average wealth or the Gini coefficient. 
For the sake of notation simplicity, from now on we assume the return process to be normally distributed and independent of time and on the value of wealth, and we denote the mean and the variance of the process as:
\begin{equation}
    \mu_r \coloneqq \mathbf{E}\left[ \hat{r}_{t}(x) \right] \qquad \sigma_r^2 \coloneqq \mathbf{Var}\left[ \hat{r}_{t}(x) \right]
\end{equation}
With this notation, condition \eqref{eq:eq_distr} and \eqref{eq:eq_average} can be written as:
\begin{equation}
\label{eq:simp_eqdist}
    \mu_r + \lim_{x\to \infty} \mathbf{E}\left[ \log\left( 1 - \frac{\hat{c}_t\left(x\right)}{x} \right) \right] < 0
\end{equation}
\begin{equation}
\label{eq:simp_eqmean}
    \left(\mu_r + \frac{\sigma_r^2}{2}\right) + \lim_{x\to \infty} \mathbf{E}\left[ \log\left( 1 - \frac{\hat{c}_t\left(x\right)}{x} \right) \right] < 0
\end{equation}
Condition \eqref{eq:simp_eqdist} and \eqref{eq:simp_eqmean} are usually assumed to be valid as a consequence of the choice of consumption function $\hat{c}_t\left(x \right)$.
Assuming a CRRA utility function\footnote{Under CRRA (Constant Relative Risk Aversion) preferences each agent is assumed to maximize the utility function $u_{\gamma}(c) = \frac{c^{1-\gamma}}{1-\gamma}$, subject to a given constraint on its wealth. This is the most commonly assumed form of utility in the economic literature on wealth inequality due to its mathematical advantages (see for example \textcite{benhabib2015wealth}).} to describe the consumption choices of the agents, we have a constant relative consumption for large values of wealth, and with the right choice of parameters the consumption can be high enough to produce a negative drift in the extreme right tail of the distribution and stabilize its evolution. 
The assumptions described in \textcite{benhabib2015wealth} are sufficient to guarantee also the existence of a finite limit for the most used inequality measures. 
We will see in the following sections, however, that for very large values of wealth the relative consumption is much smaller than the value necessary to stabilize inequality, and an additional negative drift proportional to wealth would be needed to obtain a bounded asymptotic level of inequality.

In some publications (see for example \textcite{piketty2014inequality, piketty2015wealth, gabaix2016dynamics}) wealth is normalized with the average labor income.
This is done mainly because, in order for any concept of equilibrium to make sense, the parameters of the processes defining equation \eqref{eq:each_agent} have to be constant. 
If labor incomes change with time (in distribution), they have to be normalized in order to be time-independent, and with them we are forced to normalize also wealth values. 
Rescaling wealth with any quantity increasing with time makes conditions \eqref{eq:simp_eqdist} and \eqref{eq:simp_eqmean} easier to satisfy, yet any conclusion about inequality cannot depend on the scale of the variables we are using. 
The Gini index, for example, is independent of any rescaling of wealth, hence if we find an increasing value of inequality as measured by this index, the conclusion cannot be altered by any linear change of variable. 
When passing from equation \eqref{eq:each_agent} to equation \eqref{eq:generic_growth} the same requirement of time-independence of the processes must be met in order to reach any time-independent asymptotic state.   
We obtain this condition by rescaling the variable $x_t$ by the average growth factor of labor incomes in the period under study. 
This is equivalent to consider wealth expressed in real terms instead of nominal ones, given that on average - in Italy during the period under study - labor incomes have grown in line with inflation. 
The average inflation rate over the period is $1.6\%/\text{year}$, both as declared by the Italian Institute of Statistics (\textcite{inflazione_istat}) and as derived from our data.\footnote{To derive the average growth rate of labor incomes and consumption from our survey data we studied the average shift of the distributions of the associated processes over time. 
While the shift of the averages corresponds to the inflation rate, we find that labor incomes in higher wealth classes grew faster than the ones associated to low values of wealth. 
This gradient contributes to the increase of wealth inequality, but as we will see it is not relevant to the dynamics in the right tail of the distribution, on which we will focus.} 
While too small to have an appreciable effect on labor incomes and consumption over short times, taking into consideration the real values of wealth instead of the nominal ones has immediate effects on return rates: considering real wealth is equivalent to subtracting the inflation rate from the average return $\mu_r$ expressed in nominal values. 
Thus, expressed in nominal terms, the equilibrium conditions \eqref{eq:simp_eqdist} and \eqref{eq:simp_eqmean} would become:
\begin{equation}
    \mu_r + \lim_{x\to \infty} \mathbf{E}\left[ \log\left( 1 - \frac{\hat{c}_t\left(x\right)}{x} \right) \right] < i
\end{equation}
\begin{equation}
    \left(\mu_r + \frac{\sigma_r^2}{2}\right) + \lim_{x\to \infty} \mathbf{E}\left[ \log\left( 1 - \frac{\hat{c}_t\left(x\right)}{x} \right) \right] < i
\end{equation}
where $i$ is the yearly inflation rate, $i=0.016$. 
To simplify the notation, from now on we will consider all the quantities of interest in terms of real wealth, and retain the simpler forms \eqref{eq:simp_eqdist} and \eqref{eq:simp_eqmean} for the two equilibrium conditions.
In the following sections we will assume a wealth growth process of the form \eqref{eq:generic_growth} (which, as we noted, encompasses most models studied in the economic literature), and we will use data from Bank of Italy Survey on Household Income and Wealth (SHIW) and Forbes billionaires data to infer the parameters of the processes $\hat{r}_{t}$, $\hat{c}_t(x)$ and $\hat{y}_{t}(x)$.
This estimation will provide us with a space of parameters compatible with the observed behaviour, and from this we will be able to test the validity of the conditions for the existence of an equilibrium state.

\section{Survey data}
\label{sec:shiw}
Given the general form of the evolution process \eqref{eq:generic_growth}, we have to specify the distribution of the processes $\hat{r}_{t}$, $\hat{c}_t(x)$ and $\hat{y}_{t}(x)$.
The form of the two functions $\hat{c}_t(x)$ and $\hat{y}_{t}(x)$ can be derived from Bank of Italy's Survey on Households Income and Wealth (SHIW) (\textcite{shiw2022}).
The SHIW is a survey performed by Bank of Italy from the 1960s to gather information about incomes, wealth and consumption choices of Italian households.
In its last iterations it consisted of a probabilistic sample of more than 7000 households, each declaring information about the number of its members, their incomes, consumption and wealth.\footnote{All our analysis is performed at the household level. In principle we could normalize the quantities of interest for each household by the number of its components, to obtain equivalent distributions at the individual level. The results obtained in the following would be qualitatively unchanged.
}  

We will use only a small part of the information available in the survey: in particular, we are interested in the relations between labor income, consumption and wealth. 
We want to estimate two random processes $\hat{c}_t(x)$ and $\hat{y}_{t}(x)$ in order to reproduce the empirical distributions of consumption and labor incomes found in the SHIW results.  
Aggregating all households with equal wealth value, we can study the distribution $\zeta_{x,t}(c, y)$ of the random variables $\hat{c}_t\left(x_t\right)$ and $\hat{y}_{t}\left( x_t \right)$. 
We perform the aggregation dividing the interval covered by the logarithms of wealth in the survey in 200 bins of equal amplitude. 
This implies an average standard deviation inside each bin of 0.025.
In each of the bins we evaluate the empirical (joint) distribution of consumption and labor income.
Obviously for all these distributions to be statistically significant and representative of the real underlying processes we would need a sample size much bigger than the one available from any given survey.
Luckily, as we will see, some regularities in the labor income and consumption processes allow us to describe them in a satisfactory way with the data available.

\begin{figure}[H]
    \centering
    \includegraphics[width=0.9\textwidth]{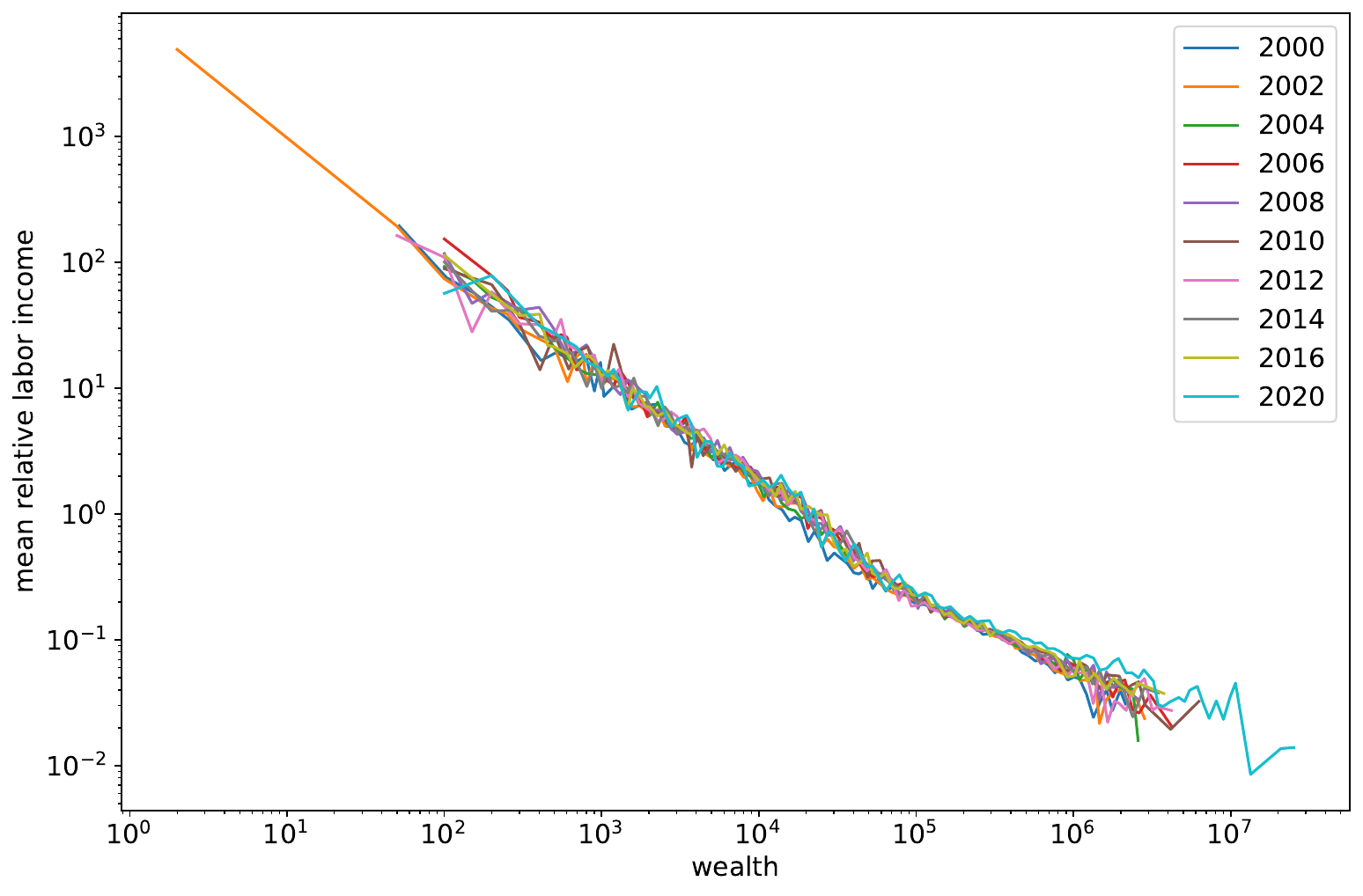}
\caption{Mean relative labor income as a function of wealth.}
\label{fig:medie_salari}
\end{figure}

We use a restricted time section of the available data: while the survey spans a range of years going from 1965 to 2020, we will limit our analysis to the period starting from 2000 and ending in 2020 (the last year available at the time of writing). 
In this time span the survey was performed 10 times, in 2000, 2002, 2004, 2006, 2008, 2010, 2012, 2014, 2016, and 2020.
The random variables we are interested in estimating from the SHIW are labor incomes and consumption relative to wealth. 
In principle we have to estimate a different random variable for each year analyzed, but we will see that in addition to regularities between different values of wealth, the variables of interest do not change significantly over the period under consideration. 
While the inflation rate has an impact over the long time evolution of the wealth distribution, as we will see the variations it produces over the time period considered are totally negligible for consumption and labor incomes. 
\begin{figure}[H]
    \centering
    \includegraphics[width=0.8\textwidth]{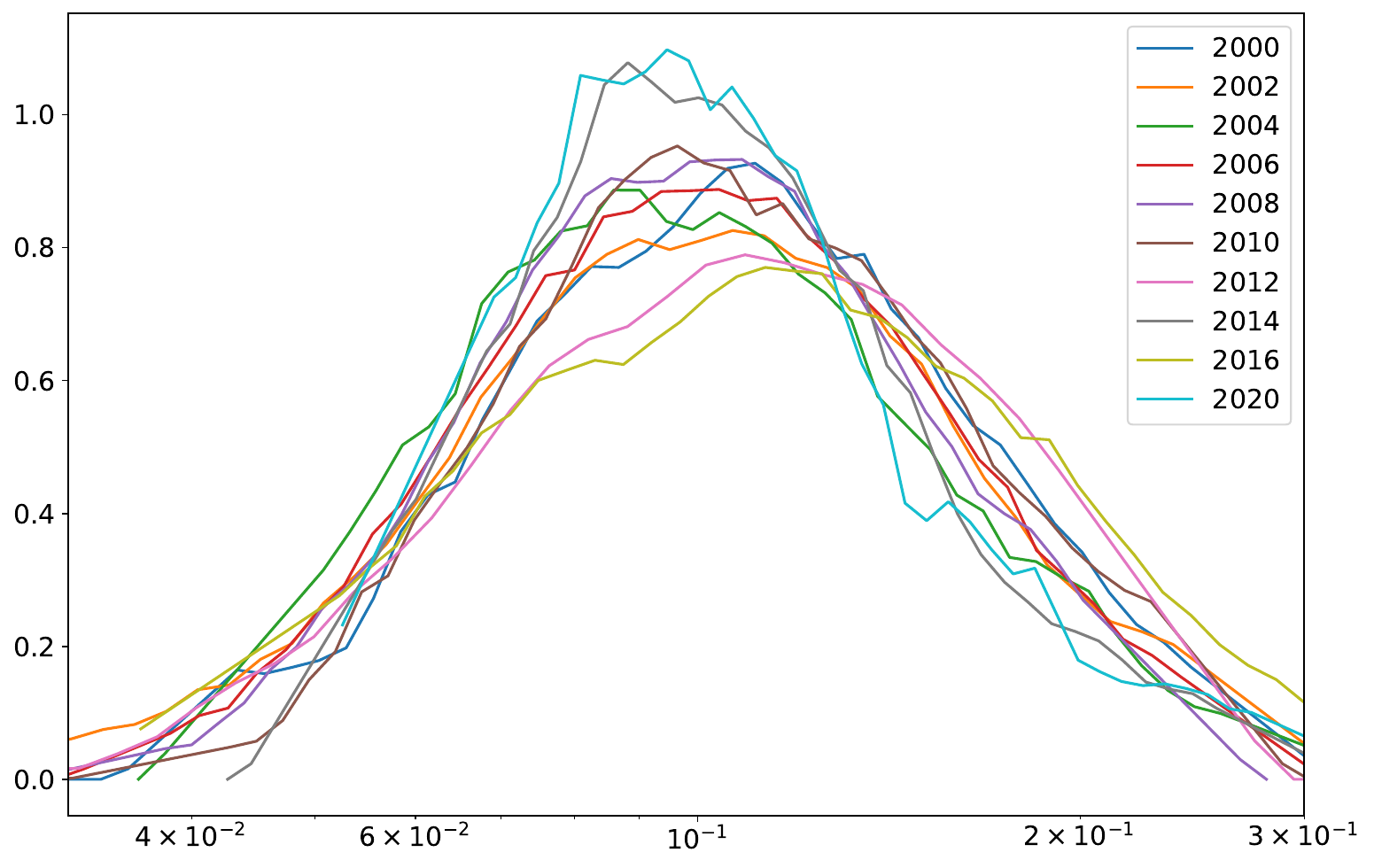}
    \caption{Distribution of relative consumption for a fixed value of wealth.}
    \label{fig:sovrapposizione_consumi}
\end{figure}
\begin{figure}[H]
    \centering
    \includegraphics[width=0.8\textwidth]{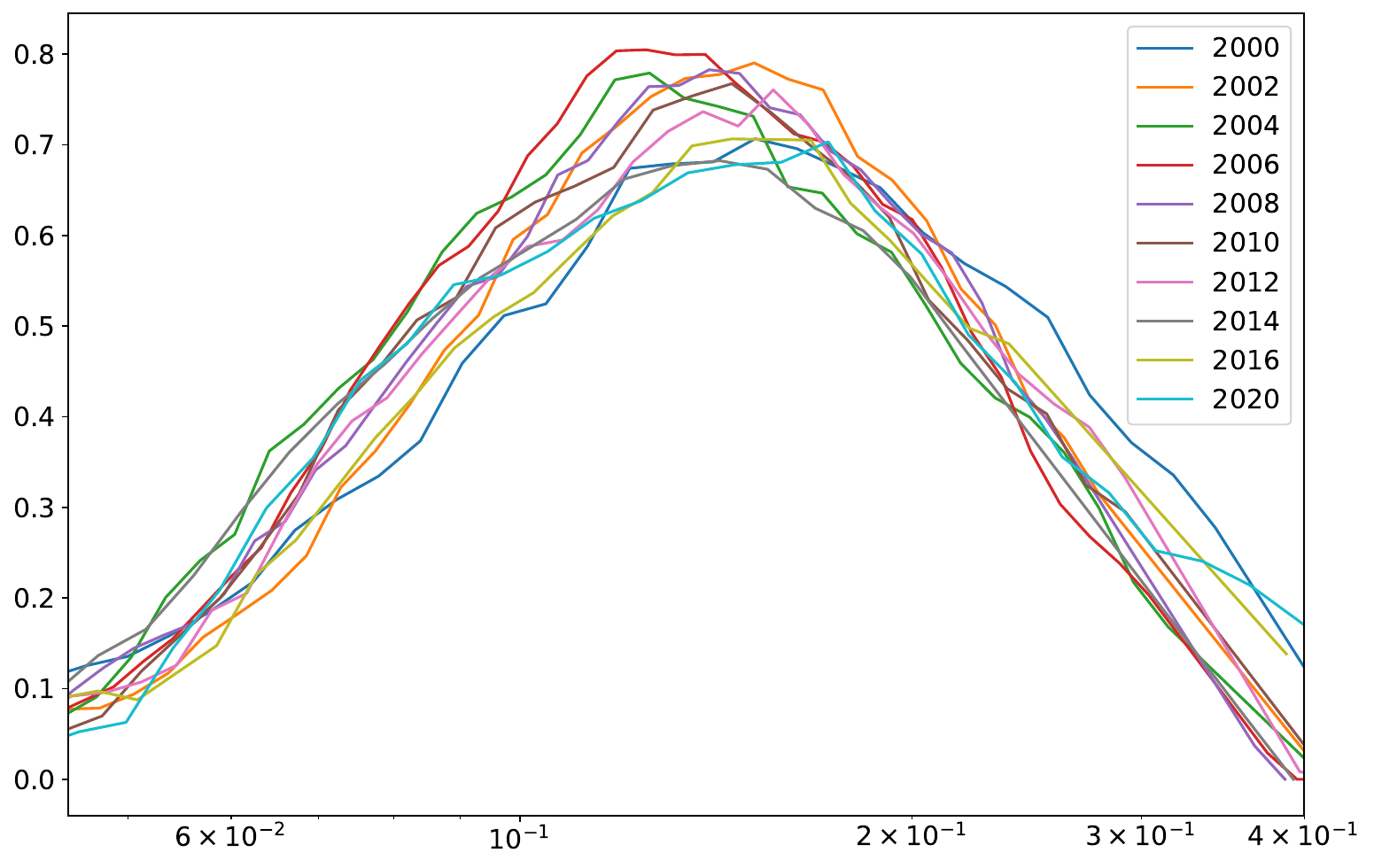}
\caption{Distribution of relative labor incomes for a fixed value of wealth.}
\label{fig:sovrapposizione_salari}
\end{figure}
In figures \ref{fig:sovrapposizione_consumi} and \ref{fig:sovrapposizione_salari} we can see the distributions of the logarithms of relative consumption and labor incomes for all the years considered, for a given wealth bin. 
Note that the variables are not standardized and are expressed in terms of nominal wealth instead of real wealth, but the distributions are nonetheless very similar throughout the period (the distributions are smoothed to make the plots clearer).
This is confirmed numerically by the variation of their first and second moments: the relative standard deviation of the means of the distributions is of $1\%$ for both the logarithms of relative consumption and the logarithms of relative labor incomes; we have standard deviations of the variances of $0.12$ and $0.06$, to be compared with average variances of $0.27$ and $0.19$ for labor incomes and consumption respectively. 
These values are in line with the standard deviations of sample variances of a normal distribution.
While the exact form of the distribution of the labor income and consumption processes is not essential for our conclusions, from figures \ref{fig:sovrapposizione_consumi} and \ref{fig:sovrapposizione_salari} and from the above numerical analysis we can see that the normality of the variations of their logarithms is a reasonable assumption.
With this assumption, we need to specify only two moments of the distributions of the logarithms for each value of wealth in order to fully identify the processes. 
Given the similarities of the labor income and consumption processes over time, for each wealth bin we take the average of the mean logarithm of relative consumption and relative labor incomes over time, and consider these averages as the theoretical means of the processes. 
We find a regular behaviour in the average (relative) labor incomes and consumption as functions of wealth, as can be seen in figure \ref{fig:medie_consumi} and \ref{fig:medie_salari}: 
\begin{figure}[H]
    \centering
    \includegraphics[width=0.9\textwidth]{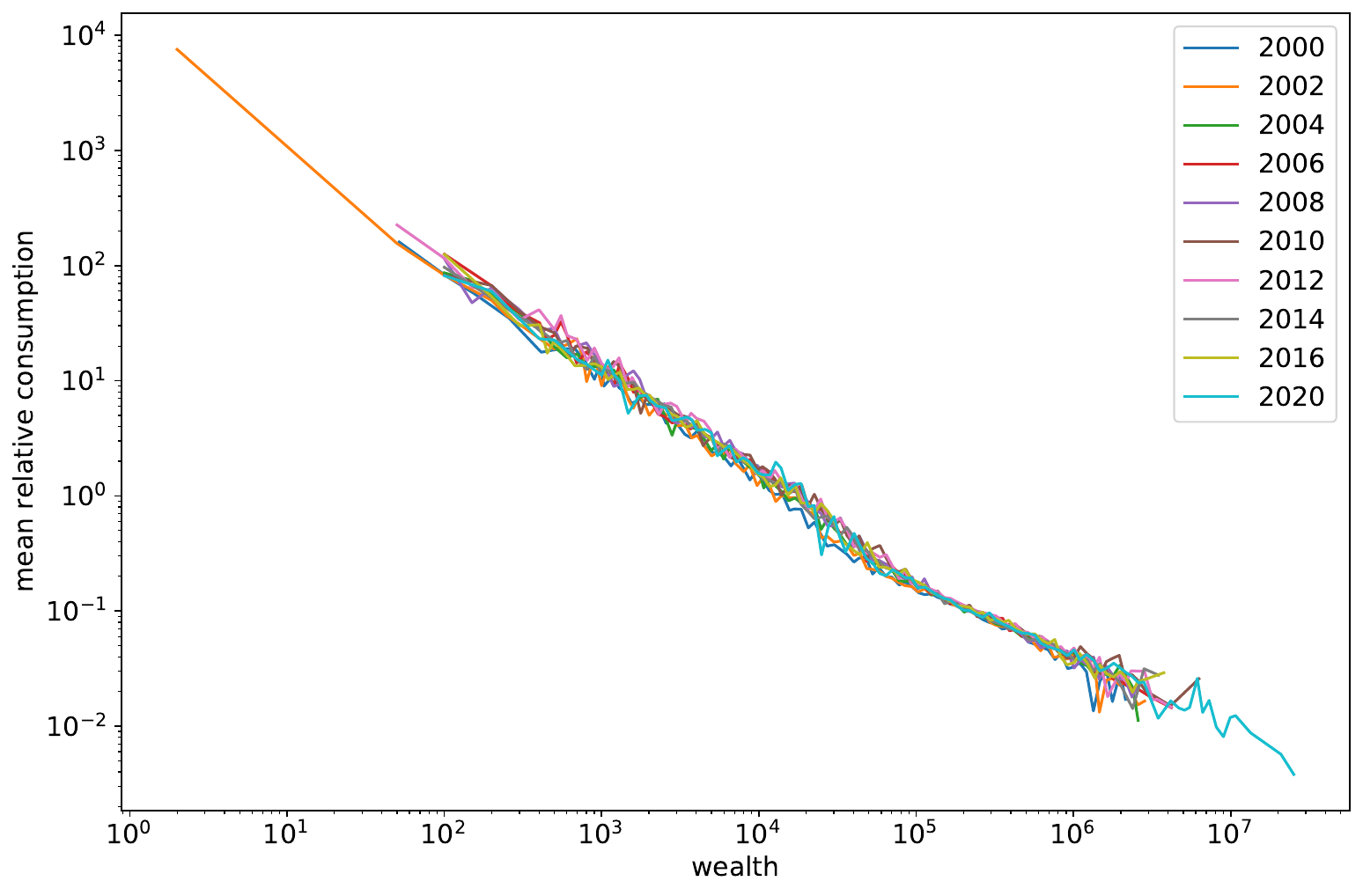}
    \caption{Mean relative consumption as a function of wealth.}
    \label{fig:medie_consumi}
\end{figure}

The relative consumption and labor incomes follow (on average) power laws as functions of wealth, with piece-wise constant exponents.
This implies that the estimation of the mean (logarithm) consumption and labor income for a given bracket of wealth does not depend only on data available in this bracket: the estimation of the mean is performed exploiting data from the whole wealth range.
The deviations of the mean (log-)relative consumption and (log-)labor income from the estimated law can be considered an effect of the small sample size in each wealth bracket. 
Most important for our analysis, we can see that both relative consumption and relative labor incomes tend to zero for large values of wealth.
Writing a generic, wealth-dependent stochastic process $\hat{z}_t(x)$ in terms of its deviations from the average logarithm $\mu_z(x)$:
\begin{equation}
    \hat{z}_t(x) = e^{\mu_z(x)} e^{\log \hat{z}_t(x) - \mu_z(x)}
\end{equation}
we can see that if the term $e^{\mu_z(x)}$ tends to zero for large values of $x$, the exact distribution of $\hat{z}_t(x)$ is not relevant as long as its dispersion is of order 1.
The same can be said for relative consumption and relative labor incomes, and that explains why the results of our analysis do not rely on the particular distribution of these processes, as long as their averages become small for large values of wealth. 

\begin{figure}[H]
    \centering
    \includegraphics[width=0.8\textwidth]{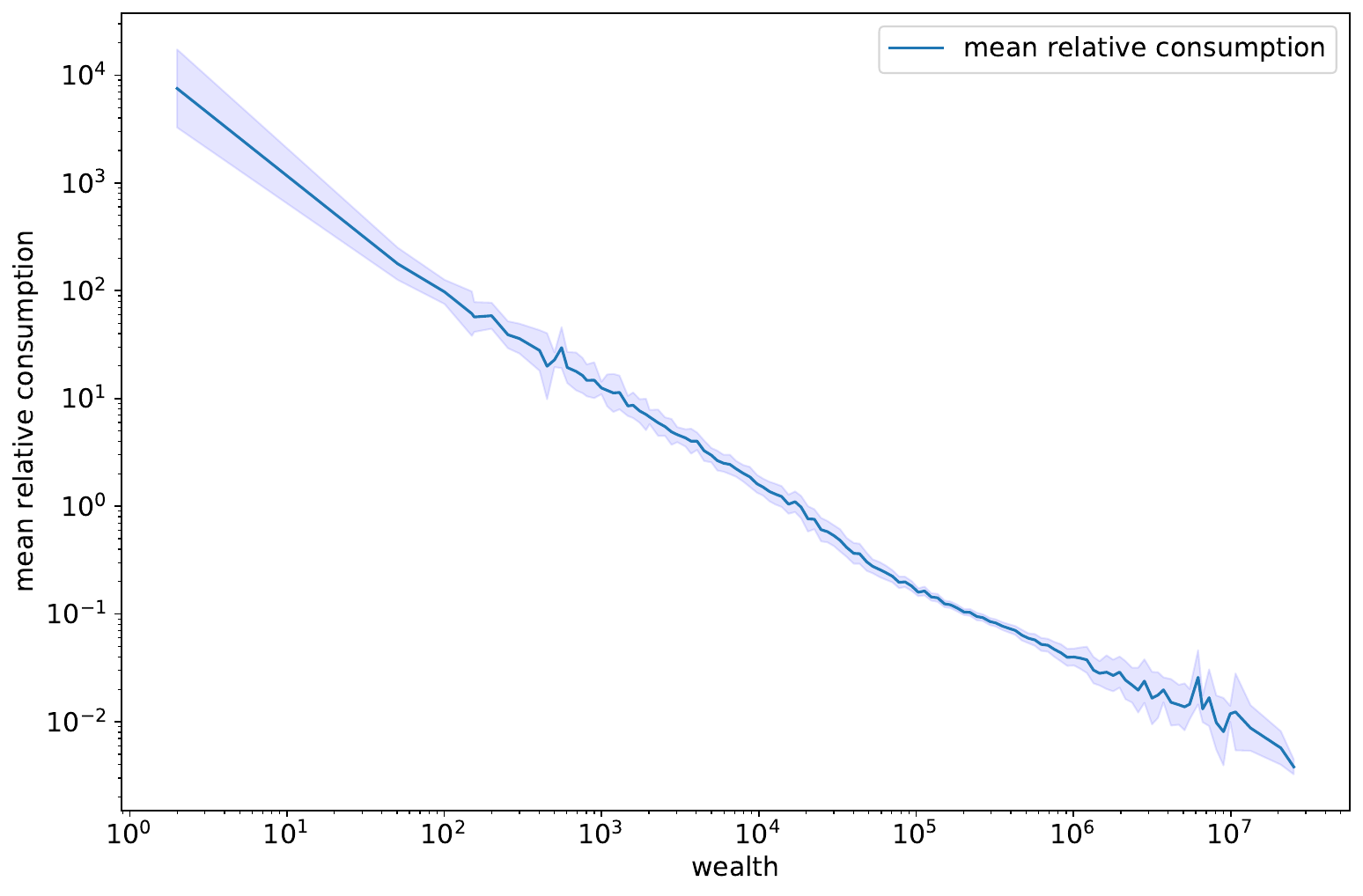}
    \caption{Time average of relative consumption as a function of wealth.}
    \label{fig:media_tempo_consumi}
\end{figure}

\begin{figure}[H]
    \centering
    \includegraphics[width=0.8\textwidth]{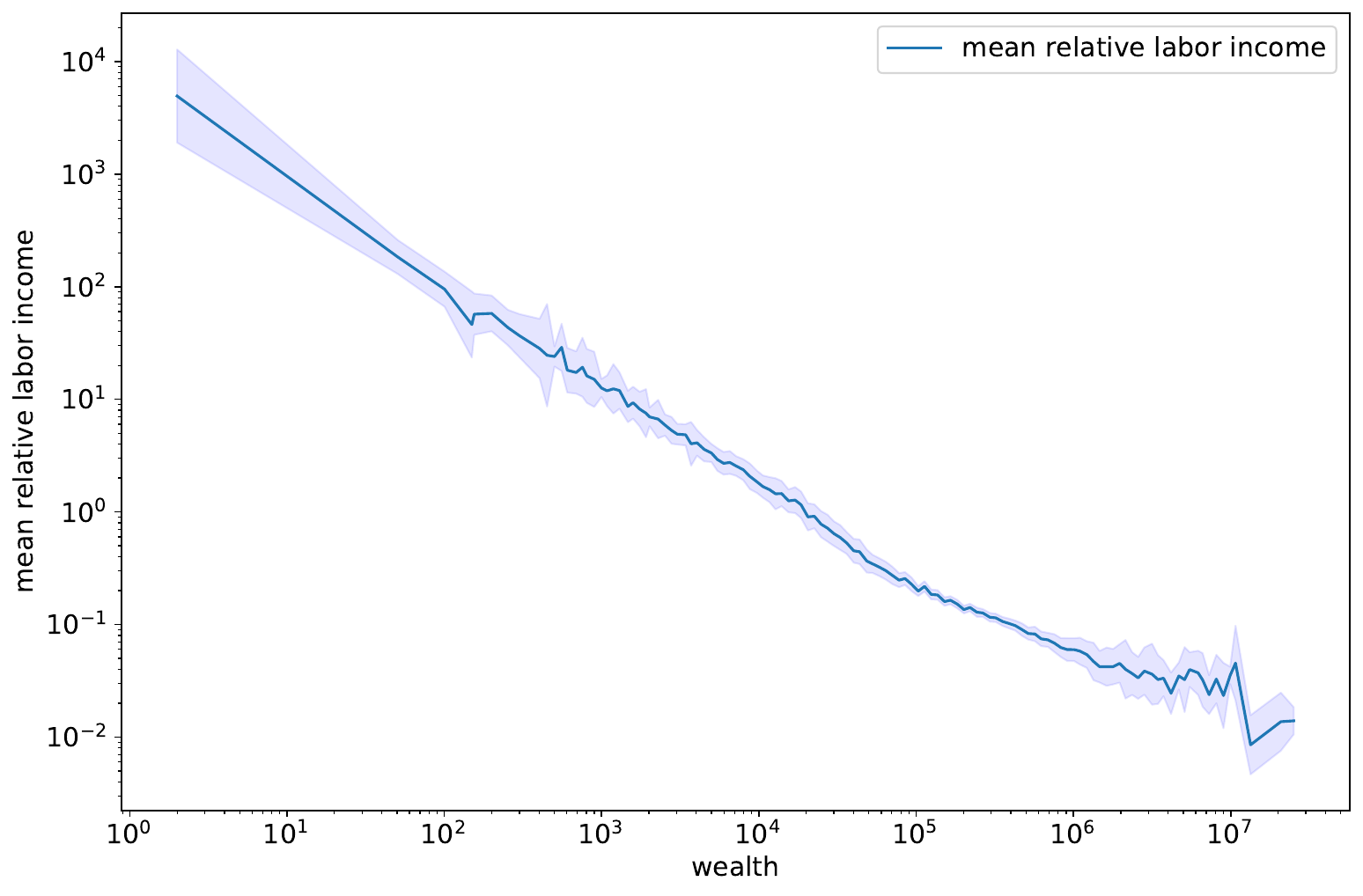}
\caption{Time average of relative labor incomes as a function of wealth.}
\label{fig:media_tempo_salari}
\end{figure}

We estimate the variances of the processes in a similar way as we did for the means, and find approximately constant values across the range of wealth analyzed, as well as over time (we are modeling the processes of \emph{relative} labor income and consumption, such that this conclusion is not really surprising, as noted in \textcite{gibrat1931inégalités}).
As already mentioned, we find for the logarithm of relative labor incomes a variance of $0.27$, while for the logarithm of relative consumption a variance of $0.19$.
Finally we have to evaluate the covariance of the two processes: as with the variances, we find an approximately constant correlation of $0.65$.
In conclusion, we describe consumption and labor incomes as a bivariate log-normal variable, with a constant covariance matrix and wealth-dependent means described by power laws with piece-wise constant exponents. 
In figures \ref{fig:media_tempo_consumi} and \ref{fig:media_tempo_salari} we plot the means of the estimated random processes with 3 standard deviation error bars.  
Extrapolating these trends out of the range of wealth analyzed, we quickly notice that the two processes cannot have the stabilizing effect they are usually assumed to have over the evolution of wealth \eqref{eq:generic_growth}. 
Extrapolating the power-law behaviour to the maximum non-empty brackets of wealth in the Italian population, we find that both labor income and consumption are completely negligible in the process of wealth growth. 
For robustness sake, in the following sections we will analyze two different scenarios: at first we will assume the trend found in the range of wealth available in the SHIW to be valid for arbitrarily large wealth; in the second scenario we will make the most conservative assumption compatible with the available data: we will assume a power-law decay for the mean relative labor income as a function of wealth, and a constant mean relative consumption for all values of wealth greater than the maximum value available in the survey.\footnote{This is compatible with the usual theoretical model of consumption in the economic literature: assuming a CRRA utility function for the agents, the relative consumption tends to a positive constant for large values of wealth.}

These data can be exploited to put some constraints on the return process in order for the equilibrium conditions \eqref{eq:simp_eqdist} and \eqref{eq:simp_eqmean} to be satisfied.
Assuming negligible consumption for the largest values of wealth we have, for returns expressed in real terms:
\begin{equation}
    \mu_r < 0
\end{equation}
\begin{equation}
    \mu_r + \frac{\sigma_r^2}{2} < 0
\end{equation}
An asymptotic relative consumption greater than zero implies a positive term on the right hand side of these inequalities. Assuming an average asymptotic relative consumption equal to the last available value in the survey, the inequalities become:
\begin{equation}
    \mu_r < 0.004
\end{equation}
\begin{equation}
    \mu_r + \frac{\sigma_r^2}{2} < 0.004
\end{equation}
In appendix \ref{app:demographic} we derive the effects of demographic factors on the wealth distribution. 
Taking these factors in consideration, the two equilibrium conditions become:
\begin{equation}
    \mu_r < 0.018
\end{equation}
\begin{equation}
    \mu_r + \frac{\sigma_r^2}{2} < 0.012
\end{equation}
Considering both a non-negligible asymptotic value of consumption and demographic effects, the equilibrium conditions are:
\begin{equation}
    \mu_r < 0.022
\end{equation}
\begin{equation}
    \mu_r + \frac{\sigma_r^2}{2} < 0.016
\end{equation}
While the demographic terms are obviously relevant, in the following we will see that estimating the return process with and without consumption we arrive to qualitatively identical conclusions.

\section{Forbes list data}
\label{sec:forbes}
In this section we present data from the Forbes billionaires lists (\textcite{forbes2022}). 
These data are obtained from \textcite{freund2016origins}, and integrated with the Forbes lists themselves. 
Forbes data have been used as a tool to study the wealth distribution in the United States in \textcite{klass2006forbes}, \textcite{korom2017enduring} and \textcite{gomez2023decomposing}.  
In \textcite{vermeulen2018fat} Forbes data are combined with traditional surveys to quantify the effects of the far right tail of the distribution on standard inequality indices, both in the US and in Europe. 
We study the returns only for billionaires present in the list at least once from 1996 to 2015.\footnote{We study the returns only \emph{forward} in time, hence we don't have any survivorship bias.} For each of these individuals, we follow the wealth as recorded in the Forbes database until 2023 (the last year available at the time of writing).   
We have 480 starting values of wealth, which are all the wealth values in the dataset excluding the last year available.
These values correspond to $53$ different households, having median starting wealth of $\$2.4$ billion and mean of $\$4.3$ billion.
We find a low (negative) correlation of $-0.07$ between returns and starting wealth, and a similarly low correlation of $-0.03$ between returns and time. 
These low values are a partial confirmation of the assumption of independence of the return process from the current wealth value and from time. 

The idea of estimating the return process from the relative variations of large values of wealth is based on the fact that, as we saw in section \ref{sec:shiw}, consumption and labor incomes become negligible in the right tail of the wealth distribution (even if a residual consumption term is assumed to remain, its magnitude is much smaller than the typical return realization).
This leaves only the return process to explain the relative change in large values of wealth. 
Even assuming average return dependent on wealth, we can see in equations \eqref{eq:eq_distr} and \eqref{eq:eq_average} that the relevant returns for the equilibrium conditions are the ones related to the right tail of the wealth distribution, hence returns derived from the Forbes lists are the meaningful sample to estimate the process.

The observed returns as extracted from the Forbes billionaires list are generated by a non-trivial process: households are present in the list only if their net wealth is above $\$1$ billion. 
This implies that if a household is present in the list in a given year, we can take into account the return on that household's wealth only if it is high enough for the household to be on the list also in the following year. 
For example we cannot observe any negative return associated to a starting wealth of $\$1$ billion. 
This could in principle introduce a positive bias in the observed returns average: the sample average cannot be a good representative of the theoretical mean of the stochastic process assumed to model returns on wealth. 
We take this problem into account with two different estimation procedures. 
In the first procedure we simulate the whole process and infer the true mean from the observed (biased) one. 
For all the available starting values of wealth we generate random returns, normally distributed with a given mean and variance, and evaluate the final wealth values attained with these returns. 
If the final wealth is still greater than $\$1$ billion, we use the associated returns for the evaluation of the sample mean.
The average \rev observed" from this process should coincide with the one observed from the real return process\footnote{In the following we perform a separate analysis taking into account also a residual consumption factor, which however does not introduce qualitative changes in the procedure.}  (this procedure constitutes an example of \emph{Approximate Bayesian Computation} technique, first described in \textcite{rubin1984bayesianly}).
To estimate a probability for the true mean and standard deviation of the return process we exploit Bayes' theorem: for true mean and standard deviation $(\mu_r, \sigma_r)$ and observed mean and standard deviation $(\mu'_r, \sigma'_r)$ we have:
\begin{equation}
    P_{\text{post}}\left( \mu_r, \sigma_r | \mu'_r, \sigma'_r \right) \propto P_{\text{prior}}\left( \mu_r, \sigma_r \right)\,W\left( \mu'_r, \sigma'_r | \mu_r, \sigma_r \right)
\end{equation}
where in addition to the prior and posterior probability densities we have the likelihood $W\left( \mu'_r, \sigma'_r | \mu_r, \sigma_r \right)$, which is given by the selection process described in the previous paragraph.
We consider a prior probability distribution uniform over the rectangle $[-0.1, 0.1]\times[0.1, 0.5]$, where the two sides are for the mean and standard deviation respectively. 
For negligible asymptotic relative consumption and labor income, we obtain a posterior distribution peaked on the mean value $0.01$ and the standard deviation value of $0.29$.
Taking into consideration an asymptotic constant relative consumption, we find a mean value for the returns of $0.013$, with standard deviation of $0.30$.
The mean return has to be higher to compensate for the (small) consumption in order to reproduce the observed variations in wealth.
The reported values can be considered in relation to the parameters of the returns distribution of the Italian stock market. Most individuals in the Italian billionaires list have a large fraction of wealth in the form of ownership of large private companies. 
If we assume these companies to have similar changes in value as the public ones, we can expect the parameters derived from our analysis to be roughly similar to the ones extracted from the stock market. 
Indeed for stocks in the FTSE Mib index the yearly return process in the last ten years had an average of $0.04$, with an average standard deviation of $0.35$.
To check the robustness of the derivation we perform the same analysis after dropping observed returns above the 90\ts{th} percentile and below the 10\ts{th} percentile of the empirical distribution. 
We obtain a higher expected value for the theoretical mean, at $0.031$, and a lower expected standard deviation, at $0.17$.
This is a reflection of the fact that the lower average return obtained with the full sample can be ascribed to negative outliers.  

The second estimation procedure we employed consisted in a standard maximum likelihood estimation of the return process given the observed data. 
Both the results of this analysis and the ones obtained from the winsorized distribution are stronger than the results reported previously, which can thus be considered to put a lower bound to the returns process mean to satisfy the equilibrium conditions.
Our estimation results for returns heterogeneity are in line with the figures obtained in the most recent empirical literature (see for example \textcite{fagereng2020heterogeneity, bach2020rich}).  
We summarize the statistics of the posterior distribution of the return process parameters, in nominal terms, in table \ref{tab:returns}.
\begin{table}[h]
    \centering
    \begin{tabular}{lcc}
        \toprule
        & Mean  & Std. Deviation \\
        \midrule
        Negligible Consumption & 0.010 & 0.29  \\
        Non-Negligible Consumption & 0.013 & 0.30  \\
        Winsorized Distribution & 0.031 & 0.17  \\
        \bottomrule
    \end{tabular}
    \caption{Summary statistics for different return distributions (nominal values).}
    \label{tab:returns}
\end{table}
The statistics are expressed in real terms in table \ref{tab:returns_real}.
\begin{table}[h]
    \centering
    \begin{tabular}{lcc}
        \toprule
        & Mean  & Std. Deviation \\
        \midrule
        Negligible Consumption & -0.006 & 0.29  \\
        Non-Negligible Consumption & -0.003 & 0.30  \\
        Winsorized Distribution & 0.015 & 0.17  \\
        \bottomrule
    \end{tabular}
    \caption{Summary statistics for different return distributions (real values).}
    \label{tab:returns_real}
\end{table}
We can see that the means of the return process are compliant with condition \eqref{eq:simp_eqdist} in both scenarios analyzed and with the winsorized distribution (taking demographic factors into account). 
The situation is radically different for condition \eqref{eq:simp_eqmean}, as we will see in the next section.

\section{Analysis of the equilibrium conditions}
\label{sec:analysis_equilibrium}

\begin{figure}[h]
    \centering
    \begin{minipage}[b]{0.45\textwidth}
        \includegraphics[width=\textwidth]{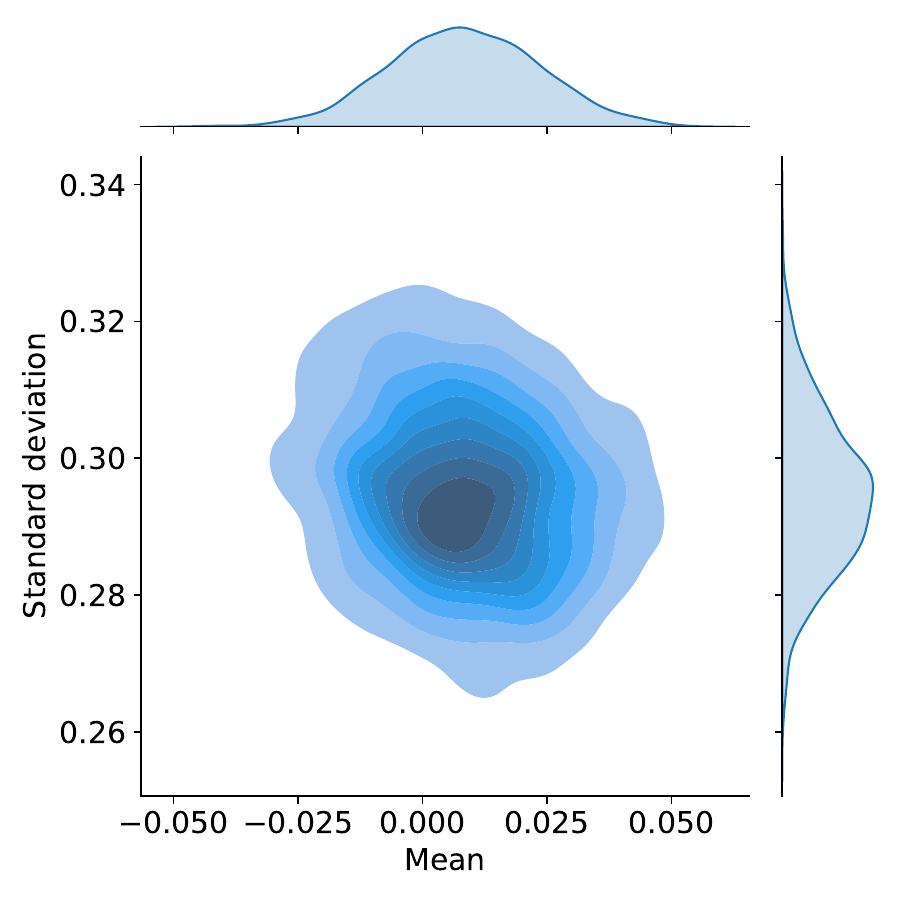}
        \caption{Posterior distribution of return process parameters with negligible consumption.}
        \label{fig:ret_dist_no_cons}
    \end{minipage}
    \hfill
    \begin{minipage}[b]{0.45\textwidth}
        \includegraphics[width=\textwidth]{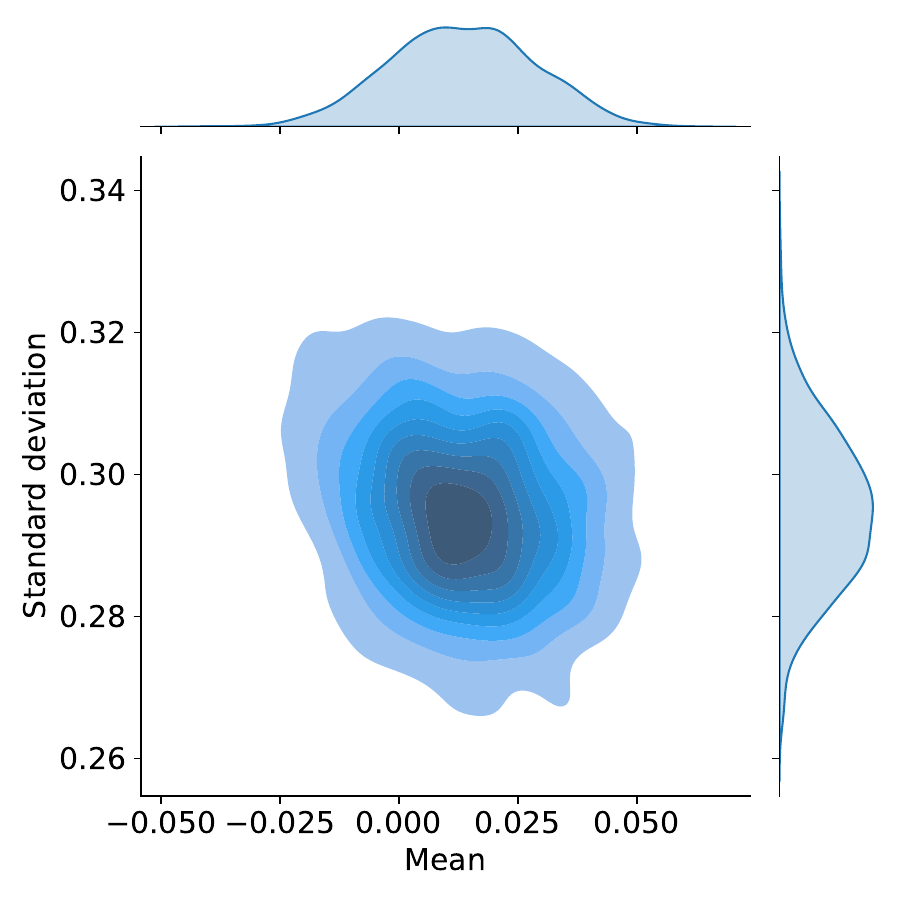}
        \caption{Posterior distribution of return process parameters with non-negligible consumption.}
        \label{fig:ret_dist_con_cons}
    \end{minipage}
\end{figure}

Beyond the expected mean and standard deviation of the return process, we obtain a full posterior distribution on the parameter space. 
With this distribution we can evaluate the likelihood to comply with conditions \eqref{eq:simp_eqdist} and \eqref{eq:simp_eqmean}. 
Both with and without the asymptotic consumption term we obtain similar results in terms of probabilities of complying with the necessary conditions for equilibrium. 
While the condition to have an equilibrium distribution is less stringent if consumption terms are taken into account, to reproduce the observed variations of wealth the average theoretical mean of the return process has to be higher, shifting the return probability distribution toward higher values.
We show in figures \ref{fig:ret_dist_no_cons} and \ref{fig:ret_dist_con_cons} the posterior distributions of the return process parameters in nominal terms, as inferred from the observed data, both with and without asymptotic consumption.

We list in table \ref{tab:soglie_prob_dist} the probabilities to have mean return below a given threshold (in real terms).
As already noticed, taking a positive asymptotic value of relative consumption into account does not change the conclusions of the analysis: the threshold for $\mu_r$ in order to attain an equilibrium distribution becomes higher, but the posterior probability distribution of $\mu_r$ is shifted toward higher values, and these two effects roughly balance out.
We obtain a probability of $56\%$ and $69\%$ to have negative mean real returns, respectively considering and neglecting consumption. 
This implies that, even ignoring demographic effects, in more than $50\%$ of the parameter space we obtain a process compatible with an equilibrium distribution. 
The results are even stronger when demographic factors are considered (see appendix \ref{app:demographic}): 
in this case we obtain that the wealth evolution process is compatible with an equilibrium distribution for more than $95\%$ of the parameter space, both with and without consumption.

\begin{table}[h]
    \centering
    \begin{tabular}{ccc}
        \toprule
        Probability of $\mu_r<\dots$ & Negligible Consumption  & Non-negligible Consumption \\
        \midrule
        0.04 & 1.00 & 1.00  \\
        0.03 & 1.00 & 0.99  \\
        \textcolor{red}{0.022} & 0.97 & \textcolor{red}{0.95}  \\
        0.02 & 0.97 & 0.93  \\
        \textcolor{red}{0.018} & \textcolor{red}{0.96} & 0.91  \\
        0.01 & 0.88 & 0.80 \\
        0.00 & 0.69 & 0.56  \\
        \bottomrule
    \end{tabular}
    \caption{Probabilities to have mean returns below given thresholds. 
    The threshold for the existence of an equilibrium distribution with negligible asymptotic consumption is $\mu_r = 0.018$; the threshold for the existence of an equilibrium distribution with non-negligible asymptotic consumption is $\mu_r = 0.022$.}
    \label{tab:soglie_prob_dist}
\end{table}
Focusing on the existence of an equilibrium value for the average wealth the situation changes dramatically. 
As we can see in table \ref{tab:soglie_prob_dist_eff}, inequality \eqref{eq:simp_eqmean} puts a much stronger constraint on the model parameters. 
In both scenarios analyzed, we find that without demographic effects an equilibrium value for the average wealth exists only for less than $1\%$ of the parameter space. 
Taking into account demographic factors the conclusion does not change qualitatively: we find that, both with and without consumption, less than $5\%$ of parameter space is compatible with the existence of an equilibrium average wealth. 
\begin{table}[h]
    \centering
    \begin{tabular}{ccc}
        \toprule
        Probability of $\left(\mu_r + \frac{\sigma_r^2}{2}\right)<\dots$ & Negligible Consumption  & Non-negligible Consumption \\
        \midrule
        0.05 & 0.83 & 0.72  \\
        0.04 & 0.61 & 0.50  \\
        0.03 & 0.35 & 0.23  \\
        0.02 & 0.14 & 0.08  \\
        \textcolor{red}{0.016} & 0.09 & \textcolor{red}{0.04}  \\
        \textcolor{red}{0.012} & \textcolor{red}{0.05} & 0.02  \\
        0.01 & 0.03 & 0.01  \\
        0.00 & 0.01 & 0.00  \\
        \bottomrule
    \end{tabular}
    \caption{Probabilities to have $\left(\mu_r + \frac{\sigma_r^2}{2}\right)$ below given thresholds. 
    The threshold for the existence of an equilibrium average wealth with negligible asymptotic consumption is $\left(\mu_r + \frac{\sigma_r^2}{2}\right)=0.012$; the threshold for the existence of an equilibrium average wealth with non-negligible asymptotic consumption is $\left(\mu_r + \frac{\sigma_r^2}{2}\right) = 0.016$.}
    \label{tab:soglie_prob_dist_eff}
\end{table}

The divergence of the average wealth is particularly significant here because it has strong implications for the asymptotic levels of wealth inequality.
In appendix \ref{app:divergence_inequality} it is shown that the divergence of the average wealth implies that also the levels of inequality are bound to diverge. 
This conclusion is not challenged by any rescaling of the wealth: as already noted the Gini index is invariant under rescaling, hence its saturation cannot be prevented with such a change of variable. 

In \textcite{berman2020wealth} the authors perform an analysis similar to ours for the economy of the United States.
They too find results incompatible with a stable equilibrium value of inequality.
Their results are in fact even stronger: over a large portion of the historical period they studied, they find a high probability for the equilibrium distribution not to exist.
In addition, when the conditions for the existence of an equilibrium distribution are respected, they evaluate the equilibration times of the wealth evolution process. They argue that for equilibration times of, say, hundreds of years, the notion of equilibrium distribution becomes meaningless in any real application, given that in the same time span many shocks would have hit the economy under study. 
Here we perform a similar exercise: we study the equilibration times of the Italian wealth distribution, assuming the conditions to attain an equilibrium distribution and a finite asymptotic average wealth are satisfied. 
In particular, we make the (rather unrealistic) assumption that there are no returns on wealth, or equivalently that the returns are null with probability 1. 
In addition we assume that the labor incomes maintain the trend observed in the available data, and  for the relative consumption to have a positive asymptotic average equal to the last value available in the survey data (realistic variations to these assumptions do not substantially weaken our conclusions). 
With these parameters we allow the evolution process to equilibrate and then, once the equilibrium distribution is attained, we perturb it increasing the wealth of a group of random households. 
We assign to these households a wealth equal to 1000 times the average value (to put this value in context, the wealthiest Italian households have wealth in the order of tens of billions of euro, while the average household wealth in Italy is of order $10^5$ euro).
In addition, to replicate the effects of demographic factors in the highest percentiles of wealth, we let the number of \rev out of equilibrium" households decay by the mortality rate in the Forbes list, $2\% / \text{year}$.
We find that, in this setting, the typical time needed for the average wealth and Gini index to come back to their equilibrium values is more than 200 years. 
Furthermore the \rev out of equilibrium" households remain on aggregate wealthier than the bulk of the population for more than 300 years. 
Without demographic effects these equilibration times are determined by the asymptotic relative consumption, and are even longer.

It is difficult to uphold the hypothesis that the process governing the evolution of wealth is described by the same parameters for such long times.
Technological and societal changes, wars and fiscal reforms must bring with them adjustments on the wealth growth dynamics, and they usually happen at a much higher frequency than what would be needed for the wealth distribution to reach its equilibrium state - assuming that this state exists.
These results are similar to the ones found in \textcite{berman2020wealth} for the United States wealth distribution. 
In \textcite{barone2016intergenerational}, the authors find a long memory in the wealth distribution: in particular, they find that family names carry information about the household wealth over a time span of 6 centuries.   
This finding is incompatible with an evolution process having equilibration times short enough for the equilibrium distribution to be relevant.
It is, on the other hand, perfectly compatible with the results described here and in \textcite{berman2020wealth}.

\section{Conclusions}
\label{sec:concl}

In this paper we presented a critical analysis of the conditions for an equilibrium treatment of the dynamics of wealth to be valid.
While the convergence to an equilibrium state is one of the most common assumptions in the economic literature on the subject, few studies focus on the necessary conditions for its applicability.
Here we used a general process for the evolution of wealth taking into account consumption, labor incomes, returns on wealth and relevant demographic factors, in order to quantify the effect of each of these elements on the stability of the wealth distribution.

We find that for large values of wealth consumption have a negligible impact, and that demographic factors are important for the stability of the distribution.
On the other hand, the greatest force pushing the distribution out of equilibrium is represented by the heterogeneity of returns on wealth, a factor often neglected in the theoretical literature (one exception can be found for example in \textcite{blanchet2022uncovering}).
While some recent studies introduced an artificial heterogeneity in the \emph{average} returns depending on wealth (\textcite{gabaix2016dynamics}), we note that this effect is in principle very different from the heterogeneity deriving from the fact that returns on wealth have almost always an unpredictable, stochastic component.
For the case of the Italian distribution of wealth, we find data compatible with the existence of an asymptotic equilibrium distribution, but not a finite asymptotic average wealth, nor a bounded level of inequality.
In other words we find that - given the properties of the dynamics of wealth observed over the last twenty years - the asymptotic distribution of wealth implies diverging inequality. 
This result is compatible with the one obtained, with different methods, for the United States wealth distribution in \textcite{berman2020wealth}.
In addition, assuming the existence of an equilibrium distribution with bounded inequality, we quantified the time needed to reach this state. 
We find equilibration times of the order of centuries, hence much longer than the typical time span between economic shocks.
This slow convergence to equilibrium supports the findings of \textcite{barone2016intergenerational}, in which the authors discover a centuries-long memory in the distributions of wealth and income.  
These results highlight the importance of the empirical assessment of some of the underlying hypotheses to the theoretical analyses of the subject. 
In particular, the finding that a weaker notion of equilibrium is valid, compared to the one usually assumed in the economic literature, underscores the necessity of incorporating out-of-equilibrium techniques to fully understand and interpret the observed phenomena. 

\section*{Acknowledgement}
The author gratefully acknowledges valuable comments from James Ridgway and Melanie Koch on a preliminary version of this draft.



\printbibliography 
 
\clearpage
 
\appendix

\section{Conditions for the existence of an equilibrium for discrete time processes}
\label{app:conditions_equilibrium}
In this appendix we show that, for evolution described by \eqref{eq:generic_growth}, a necessary condition for the existence of an equilibrium distribution is given by:
\begin{equation}
    \lim_{x\to \infty} \mathbf{E}\left[ \hat{r}_{t}(x) + \log\left( 1 - \frac{\hat{c}_t\left(x\right)}{x} \right) \right] < 0 
\end{equation}
For large values of wealth $x_t$ we saw in section \ref{sec:shiw} that the labor income contribution becomes negligible, and relative consumption tends to a constant (possibly zero). In this setting, equation \eqref{eq:generic_growth} can be approximated by:
\begin{equation}
\label{eq:reduced_growth}
    \hat{x}_{t}\left(x_t\right) = x_t\,e^{\hat{r}_{t}}\left( 1 - \hat{c}_{t} \right)
\end{equation}
In this form the evolution of wealth is a pure multiplicative random process, and can be expressed as:
\begin{equation}
\label{eq:wealth_random_walk}
    \log \hat{x}_{t}\left(x_t\right) = \log x_t + \hat{\ell}_{t}
\end{equation}
with
\begin{equation}
    \hat{\ell}_{t} = \hat{r}_{t} + \log \left( 1 - \hat{c}_{t} \right) \qquad  \hat{c}_{t} = \lim_{x \to \infty} \frac{\hat{c}_t\left(x\right)}{x}
\end{equation}
Equation \eqref{eq:wealth_random_walk} describes a random walk for the logarithm of wealth, and from the properties of this random walk we can derive the shape of the (right tail of the) long term distribution of wealth. 
 
We pick a threshold wealth $x_*$, above which equation \eqref{eq:reduced_growth} is a good enough approximation to \eqref{eq:generic_growth}. If an equilibrium distribution exists, there can be no probability flux across this threshold:
\begin{equation}
    \frac{\partial P\left(x_t>x_* \right)}{\partial t} = 0 
\end{equation}
This implies the threshold can be treated as a reflecting barrier for the process \eqref{eq:wealth_random_walk}, and the results in \textcite{levy1996power, sornette1997convergent} applies. 
There the authors prove that, for a random walk with a reflecting barrier, the long term distribution of the position is described by an exponential distribution:
\begin{equation}
\label{eq:exponential_dist}
    \rho(z_t) \propto e^{-\mu\,z_t}
\end{equation}
where $z_t$ is the position of the random walker (in our case $z_t = \log x_t$), and $\mu$ is given by the equation:
\begin{equation}
\label{eq: defining_condition}
    \int d\ell\, w(\ell) \, e^{\mu \, \ell} = 1
\end{equation}
with $w(\ell)$ the distribution of the random walk steps defined in \eqref{eq:wealth_random_walk}. 
Exploiting the normal distribution of $\hat{r}_t$ and the fact that $\mathbf{E}\left[ \hat{c}_{t}\right] \ll 1$, we find:\footnote{The normality assumption is not strictly necessary to derive equation \eqref{eq:mu_def}: the integral in equation \eqref{eq: defining_condition} can be evaluated perturbatively or numerically with the empirical distribution of returns $\hat{r}_t$, to verify that the corrections generated by non-gaussian terms are negligible.}
\begin{equation}
\label{eq:mu_def}
    \mu = - \frac{2\, \mathbf{E} \left[ \hat{\ell_t}\right]}{\mathbf{Var} \left[ \hat{\ell_t}\right]}
\end{equation}
Expressing the exponential distribution \eqref{eq:exponential_dist} in terms of $x_t$, we find:
\begin{equation}
    \rho(x_t) \propto x_t^{-\left( 1 + \mu \right)}
\end{equation}
and as anticipated, for an equilibrium distribution to exist we must have:
\begin{equation}
    \mu > 0 \Longrightarrow \mathbf{E} \left[ \hat{\ell_t}\right] = \mathbf{E} \left[ \hat{r}_{t}\right] + \mathbf{E} \left[ \log\left( 1 - \hat{c}_t \right)\right] < 0
\end{equation}
In addition, for the average to exist we must have $\mu > 1$, which translates to:
\begin{equation}
    \mathbf{E} \left[ \hat{\ell_t}\right] < -\frac{\mathbf{Var} \left[ \hat{\ell_t}\right]}{2}
\end{equation}
Expressing the inequality in terms of $\hat{r}_{t}$ and $\hat{c}_t$ and neglecting the variance of $\hat{c}_t$, this can be written as:
\begin{equation}
\label{eq: av_existence_condition}
    \mathbf{E} \left[ \hat{r}_{t}\right] + \mathbf{E} \left[ \log \left( 1 - \hat{c}_t \right)\right]  + \frac{\mathbf{Var} \left[ \hat{r}_{t}\right]}{2} < 0
\end{equation}
Finally, if an asymptotic value for the average wealth does not exist we prove in appendix \ref{app:divergence_inequality} the divergence of inequality in the wealth distribution, such that equation \eqref{eq: av_existence_condition} represents also a necessary condition to have a finite asymptotic inequality.

\section{Divergence of inequality}
\label{app:divergence_inequality}

Here we want to prove that - assuming wealth dynamics governed by equation \eqref{eq:generic_growth} and labor income and consumption process as described in section \ref{sec:shiw} - a diverging value of the average wealth implies a saturating Gini index. 
If inequality \eqref{eq:eq_distr} is not satisfied, i.e. no equilibrium distribution exists, asymptotically the dynamics of all agents will be driven by the simplified equation \eqref{eq:reduced_growth}. 
Indeed, above the wealth threshold $x_*$ the dynamics has a positive drift, such that each agent with wealth above the threshold will tend to acquire an even bigger wealth (and there is always a finite probability to move from below to above the threshold, so the probability for each agent to be found below the threshold tends to zero with time).
If, on the other hand, inequality \eqref{eq:eq_distr} is satisfied, an equilibrium distribution exists, and the probability to find an agent with wealth greater than any given value tends to a constant value. 
If the average wealth $\mathbf{E}\left[ x_t \right]$ diverges with time, the probability to have wealth greater than a fixed multiple of the average has a vanishing limit:
\begin{equation}
    \forall \kappa > 0 \quad \lim_{t \to \infty} \mathbf{P}\left( x_{t} \geq \kappa\, \textbf{E}\left[ x_{t} \right] \right) = 0
\end{equation}
Denoting with $p_t(x)$ the distribution of wealth at time $t$, for the Gini index $G_t$ we have:
\begin{align}
    G_t & = \frac{1}{2} \int dx dy\, p_{t}(x)\, 
    p_{t}(y) \frac{|x-y|}{\textbf{E}\left[ x_{t} \right]} = \\
    & = \int_{x\geq y} dx\, dy\, p_t(x)\, p_t(y) \frac{\left(x-y\right)}{\textbf{E}\left[ x_{t} \right]} \geq \\
    & \geq \int_{\kappa \textbf{E}\left[ x_{t} \right]}^{\infty} dx \, p_t(x) \int_{0}^{\kappa \textbf{E}\left[ x_{t} \right]} dy\, p_t(y) \frac{\left(x-y\right)}{\textbf{E}\left[ x_{t} \right]}  \geq \\
    & \geq P_t\left(x <\kappa\, \textbf{E}\left[ x_{t} \right]  \right) \int_{\kappa \textbf{E}\left[ x_{t} \right]}^{\infty} dx \, p_t(x) \left(\frac{x}{\textbf{E}\left[ x_{t} \right]} -\kappa \right)  \geq \\
    & \geq  P_t\left(x <\kappa\, \textbf{E}\left[ x_{t} \right]  \right) \left( 1 - \kappa \right)
\end{align}
where in the last inequality we used the bound:
\begin{equation}
    \int_{\kappa \textbf{E}\left[ x_{t} \right]}^{\infty} dx\, p_t(x)\, x = \textbf{E}\left[ x_{t} \right]  - \int_{0}^{\kappa \textbf{E}\left[ x_{t} \right]} dx\, p_t(x)\, x \geq \textbf{E}\left[ x_{t} \right] \left( 1 - \kappa\,\mathbf{P}\left(x_t <\kappa\, \textbf{E}\left[ x_{t} \right]  \right) \right) 
\end{equation}
The last result shows that a vanishing $\mathbf{P}\left( x_{t} \geq \kappa\, \textbf{E}\left[ x_t \right] \right)$ implies $G_t \to (1 - \kappa)$ for arbitrarily small $\kappa$, i.e. a saturation of the Gini index.

\section{Evolution process in continuous time}
\label{app:general_continuous}

In general, a stochastic evolution process for wealth can be written as:
\begin{equation}
    x_{t+1} = x_{t} + \Delta x_t\left( x_{t} \right)
\end{equation}
When the variation $\Delta x_t\left( x_{t} \right)$ is small with high probability\footnote{More precisely: if for any function $f\left(x\right)$ we are interested in evaluating we have $\left(\frac{d\log f(x_t)}{d x_t}\right)\Delta x_t\left( x_{t} \right) \ll 1$ almost surely. This condition can be attained for any infinitely-divisible growth process by considering small enough time-steps $\Delta t$.}, we can approximate the evolution with a continuous process, and obtain the equation:
\begin{equation}
\label{eq: kolmogorov}
    \frac{\partial \rho_t(x)}{\partial t} = - \frac{\partial }{\partial x} \left( \rho_t(x) A(x) \right) + \frac{1}{2} \frac{\partial^2 }{\partial x^2}\left( \rho_t(x) B(x) \right)
\end{equation}
where $\rho_t(x)$ is the probability density to find an agent with wealth $x$, and:
\begin{equation}
    A(x) =  \mathbf{E}\left[ \Delta x_t\left( x_t \right) | x_t=x \right]
\end{equation}
\begin{equation}
    B(x) =  \mathbf{E}\left[ \Delta x_t\left( x_t \right)^2 | x_t=x \right] \approx \mathbf{Var}\left[ \Delta x_t\left( x_t \right) | x_t=x \right]
\end{equation}
Where in the last equation we assumed $A(x)^2$ to be negligible.
Integrating equation \eqref{eq: kolmogorov} from some value $x$ to infinity, we obtain:
\begin{equation}
\label{eq:integrated_kolmogorov}
    \frac{\partial P_t(x_t > x )}{\partial t} = \rho_t(x) A(x)  - \frac{1}{2} \frac{\partial }{\partial x}\left( \rho_t(x) B(x) \right)
\end{equation}
For high enough values of $x$, we can make some assumptions on $A(x)$ and $B(x)$. In particular we can assume the variation of wealth to be proportional to the wealth itself\footnote{The evolution of wealth can be described, for large values of wealth, by a multiplicative process: $ x_{t+1} = e^{r_{t}} (1-c_t) x_{t}$. Hence we have $\Delta x_t\left( x_t \right) = \left(e^{r_{t}} - 1\right)(1-c_t) x_{t}$. For negligible relative consumption we obtain the coefficients: $\alpha \approx \mu_r + \frac{\sigma_r^2}{2}$, $\beta \approx \sigma_r^2$, with $\mu_r$ and $\sigma_r^2$ the mean and variance of the process $r_{t}$.}:
\begin{equation}
    A(x) = \alpha\, x \quad , \quad B(x) = \beta\, x^2
\end{equation}
From this, we can prove that the tail of the equilibrium distribution of \eqref{eq:integrated_kolmogorov} can be described by a power law.
Indeed, equation \eqref{eq:integrated_kolmogorov} becomes in equilibrium:
\begin{equation}
    \rho_t(x)\, x^2  = \frac{\beta}{2\, \alpha} \frac{\partial }{\partial \log x}\left( \rho_t(x)\, x^2 \right)
\end{equation}
Rearranging the terms we obtain:
\begin{equation}
    \frac{\partial \log \left( \rho_t(x) x^2 \right) }{\partial \log x} = \frac{2\,\alpha}{\beta}
\end{equation}
The solution of the last equation is given by:
\begin{equation}
    \rho_t(x) = \left(\frac{x_0}{x}\right)^{\gamma}
\end{equation}
for some $x_0$ and $\gamma=2\left(1 - \frac{\alpha}{\beta} \right)$.
In terms of the discrete process, this becomes:
\begin{equation}
\label{eq:exponent}
    \gamma = 1 - \frac{2\,\mu_r}{\sigma_r^2}.
\end{equation}

\section{Demographic terms}
\label{app:demographic}

To take into account demographic effects in the evolution of the distribution of wealth we have to pass to the continuous-time limit of the evolution process, described in appendix \ref{app:general_continuous}.
We assume the total number of agents in our population to be constant in time: for each agent who disappears from the population, a new one is born. 
All the newborn agents are assumed to have zero wealth. 
The variation of the distribution of wealth generated by the death-birth process can be written as:
\begin{equation}
    \frac{\partial \rho_t(x)}{\partial t} = b_t \,f_b(x) - d_t(x)\, \rho_t(x)
\end{equation}
where $\rho_t(x)$ is the wealth distribution at time $t$, $b_t$ is the rate of birth, $d_t(x)$ is the average death rate for a given wealth value, and $f_b(x)$ is the wealth distribution of newborns, concentrated around zero wealth.
In order for the number of agents to be constant, we must have that the birth rate is equal to the average rate of death:
\begin{equation}
    b_t = \int dx\,d_t(x)\, \rho_t(x)
\end{equation}
Given that the starting wealth of newborn agents is concentrated around zero, births have no bearing on the evolution of the right tail of the distribution. 

In addition to births and deaths, we have to model inheritances. 
For each dying agent with wealth $x$, we assume its wealth to be distributed equally among $j$ heirs, where $j$ is extracted from a Poisson distribution.
To fix the parameter of the distribution we would need the average number of inheritors per succession. 
While this number is not directly available for the Italian population, in \textcite{elinder2018inheritance} the authors estimate the average number of heirs in Sweden to be approximately $2.9$. 
We assume this value to describe in a satisfactory way also the Italian population, also considering that the precise value of this parameter does not alter our final conclusions. 
Finally, we are interested in the inheritance from agents in the extreme right tail of the distribution, hence we can assume that the wealth of the heirs before the inheritance is negligible with respect to the inherited wealth. 
With all this, the time variation of the distribution of wealth caused by demographic changes and inheritances is given by:
\begin{equation}
    \frac{\partial \rho_t(x)}{\partial t} = b_t \,f_b(x) - d_t(x)\, \rho_t(x) + \sum_{j=1}^{\infty} p_j\, j\,  d_t\left(j\,x\right) \rho_t\left( j\,x\right)  
\end{equation}
where $p_j$ is the probability to have $j$ inheritors.
Given that we are interested in changes in the right tail of the distribution, we can take the limit of this equation for very large values of $x$, and use the Italian billionaires list to estimate its free parameters. 
The average yearly death rate\footnote{To evaluate this average, we consider the age of each member of the billionaires list in a given year, associate to it the national probability of death for that age (\textcite{mortalita_istat}), and then average these values.} for Italian Billionaires in the period under study is $d = 2\%/\text{year}$.
For large values of wealth, we obtain:
\begin{equation}
    \frac{\partial \rho_t(x)}{\partial t} \approx  d \left[\sum_{j=1}^{\infty} p_j\, j\, \rho_t\left( j\,x\right) - \rho_t\left( x\right) \right]
\end{equation}
In addition we can exploit the asymptotic properties of the distribution of wealth derived in appendices \ref{app:conditions_equilibrium} and \ref{app:general_continuous}: in particular the fact that the right tail of the distribution will be asymptotically described by a power law of exponent $\gamma$. 
With this, we can factorize the term $\rho_t\left( j\,x\right)$ to obtain:
\begin{eqnarray}
        \frac{\partial \rho_t(x)}{\partial t} & \approx &  d  \left[\sum_{j=1}^{\infty} p_j\, j^{1-\gamma} - 1 \right] \rho_t\left( x\right) = \\
        & = & d \left( \mathbf{E}\left[ j^{1-\gamma}\right] - 1 \right) \rho_t\left( x\right)
\end{eqnarray}
where we defined the average $\mathbf{E}\left[ j^{1-\gamma}\right] = \sum_{j=1}^{\infty} p_j\, j^{1-\gamma}$.
Combining the effects on the wealth distribution of the mortality rates with the other processes we obtain:
\begin{equation}
        \frac{\partial \rho_t(x)}{\partial t} = - \frac{\partial }{\partial x} \left( \rho_t(x) A(x) \right) + \frac{1}{2} \frac{\partial^2 }{\partial x^2}\left( \rho_t(x) B(x) \right) + d \left( \mathbf{E}\left[ j^{1-\gamma}\right] - 1 \right) \rho_t(x)
\end{equation}
where $A(x)$ and $B(x)$ were defined in appendix \ref{app:general_continuous}.
At equilibrium we obtain:
\begin{equation}
   - \left(\mu_r + \frac{\sigma_r^2}{2} \right)  \left( 1-\gamma \right) + \frac{\sigma_r^2}{2} \left(2 - \gamma \right)\left(1 - \gamma \right) + d \left( \mathbf{E}\left[ j^{1-\gamma}\right] - 1 \right) = 0
\end{equation}
This equation can be simplified to:
\begin{equation}
\label{eq:sol_gamma}
    \left[ \mu_r  + \frac{\sigma_r^2}{2} \left( \gamma -1 \right) \right] = d \,g\left(\gamma  \right) 
\end{equation}
where we defined the function: 
\begin{equation}
    g\left( \gamma\right) = \frac{ \mathbf{E}\left[ j^{1-\gamma}\right] - 1}{\left( 1-\gamma \right)}
\end{equation}
For $d=0$ this equation reduces to \eqref{eq:exponent}. The function $g\left( \gamma\right)$ is decreasing in $\gamma$, going from $g\left( \gamma\right) = \mathbf{E}\left[ \log j \right]$ when $\gamma = 1$ to $g\left( \gamma\right) =0$ for $\gamma$ going to infinity.
The left hand side of the equation, on the other hand, is an increasing function of $\gamma$, going from $\mu_r$ for $\gamma = 1$ to infinity for diverging $\gamma$.
Hence equation \eqref{eq:sol_gamma} has a solution with $\gamma > 1$ if and only if: 
\begin{equation}
    \mu_r < \mathbf{E}\left[ \log j \right]\,d
\end{equation}
Substituting the parameter of the distribution $p_j$ and the value of the mortality rate $d$, we obtain:
\begin{equation}
    \mu_r < 0.018
\end{equation}
Evaluating the same condition with an average number of inheritors $\lambda = 2$ and $\lambda =4$ we obtain, respectively:
\begin{equation}
    \mu_r < 0.012
\end{equation}
\begin{equation}
    \mu_r < 0.025
\end{equation}
so we see that the condition has only a weak dependence on the assumptions about the average number of inheritors. 
If we want $\gamma>2$, in order to have a finite equilibrium average wealth, the condition for the existence of a solution becomes: 
\begin{equation}
        \mu_r  + \frac{\sigma_r^2}{2} < \left( 1 - \mathbf{E}\left[ j^{-1} \right] \right) d 
\end{equation}
Substituting the parameter of the distribution $p_j$ and the value of the mortality rate $d$, we find that an equilibrium value of the average wealth exists only if:
\begin{equation}
    \mu_r  + \frac{\sigma_r^2}{2} < 0.012
\end{equation}
Replicating the same calculation but assuming an average number of inheritors $\lambda = 2$ and $\lambda =4$ we obtain, respectively:
\begin{equation}
    \mu_r  + \frac{\sigma_r^2}{2} < 0.010
\end{equation}
\begin{equation}
    \mu_r  + \frac{\sigma_r^2}{2} < 0.014
\end{equation}
Hence we see that also this condition depends weakly on the assumption about the average number of inheritors.

\end{document}